\documentclass[twocolumn,showpacs,preprintnumbers,amsmath,amssymb]{revtex4}

\usepackage{graphicx}
\usepackage{dcolumn}
\usepackage{bm}

\begin{document}


\title{Improved estimation of Fokker-Planck equations through optimisation}

\author{A. P. Nawroth}
\author{J. Peinke}
\affiliation{%
Institut for Physics, Carl-von-Ossietzky University Oldenburg, D-26111 Oldenburg, Germany
}%
\author{D. Kleinhans}
\author{R. Friedrich}
\affiliation{%
Institut for Theoretical Physics, University of M\"unster, D-48149 M\"unster, Germany
}%

\date{\today}

\begin{abstract}
An improved method for the description of hierarchical complex systems by means of a Fokker-Planck equation is presented. In particular the limited-memory Broyden-Fletcher-Goldfarb-Shanno algorithm for constraint problems (L-BFGS-B) is used to minimize the distance between the numerical solutions of the Fokker-Planck equation and the empirical probability density functions and thus to estimate properly the drift and diffusion term of the Fokker-Planck equation. The optimisation routine is applied to a time series of velocity measurements obtained from a turbulent helium gas jet in order to demonstrate the benefits and to quantify the improvements of this new optimisation routine.
\end{abstract}

\maketitle

\section{\label{introduction}Introduction}

Most complex systems can be assigned to the two following classes, the time dependent complex systems and the scale dependent complex systems \cite{peinke2002}. Examples for the first class are nonlinear chaotic dynamical systems, while systems with a scaling behaviour over a wide range of different scales, like turbulence, financial markets or earth quakes are examples for the second class. Besides the characterisation of new features of these complex systems it is a challange to derive effective underlying equations for their description. A successful approach to such systems is a description through stochastic equations (Langevin or corresponding Fokker-Planck equations) which may involve nonlinearity in the deterministic as well as in the stochastic part. This approach has become particulary interesting, as it has been shown that it is possible to estimate the underlying stochastic equations directly by data analysis.

The verification of the preconditions and the application of this approach to time dependent systems has been described in \cite{siegert98,friedrich98,friedrich00,siefert03, waechter2004b}. It was successfuly applied to the description of noisy electrical circuits \cite{friedrich00}, systems with feedback delay \cite{frank2004}, traffic flow data \cite{kriso02} and physiological time series \cite{kuusela04, ghasemi2006}, to mention just a few. Also the second class, the scale dependent complex systems, which in general are not stationary in scale, can be analysed succesfully by this approach. In this class, stochastic processes evolving in scale are reconstructed. A complete statistical description, i.e. general n scale joint statistics, for certain classes of systems, such as the roughness of surfaces \cite{waechter03,jafari2003}, turbulence \cite{friedrich97b, renner01, tutkun2004} and finance \cite{friedrich00b, renner01b,ausloos2003} can be obtained. Though in general  a reconstruction of time series for the scale dependent complex systems is not possible in such a simple way as for the first class, certain promising attempts have been made \cite{jafari2003, nawroth2006a}.

The use of Langevin and Fokker-Planck equations is therefore a very promising method for time series analysis. The critical part in this method is the correct estimation of the coefficients of the Langevin or the corresponding Fokker-Planck equation, which are the so called Kramers-Moyal coefficients. A correct estimation of these coefficients is crucial to a good description of the underlying processes. The estimation of the Kramers-Moyal coefficients is complicated by the fact, that the approach itself is based on the assumption of Markov properties. This assumption is valid for many systems for big and small but finite timesteps. The main difficulties arise from the fact that for the estimation of the underlying equations it is necessary to calculate the limit of infinitely small time steps, where the Markov properties are often no longer valid. For more details concerning this discussion see \cite{ragwitz01,friedrich02,ragwitz02b}. Further concerns about systematic estimation problems were discussed in \cite{sura02}. Due to these problems in the estimation process it was necessary till now to apply manual corrections \cite{renner01b} to the determined Kramers-Moyal coefficients in some cases in order to get an optimal description. For time dependent systems these  problems were adressed in \cite{kleinhans2005} by proposing an improved estimation method for the necessary parameters. This improved estimation method utilises the comparison of the probability density functions (pdfs) generated by the Langevin equation and those computed directly from the empirical data.

In this work we address the crucial problem of the correct estimation of the coefficients for the Fokker-Planck equation for the second class of systems with scale dependent complexity. It should be noted, that there is no principle problem to transfer the results of this work to the class of time dependent complex systems. Furthermore, the methods proposed in this work can be regarded as a systematic way to include the manual corrections described for example in \cite{renner01b}. In general for the class of scale dependent complex systems the processes are not stationary in scale and so it is often necessary to use numerical instead of analytic solutions. Therefore the knowledge of the solution to the Fokker-Planck equation will not be global but point wise in the space spanned by the coefficients of the Fokker-Planck equation. In order to utilise the comparison of the pdfs, as it has been suggested for the first class of complex systems \cite{kleinhans2005}, optimization routines are proposed to find the optimal set of parameters, which implies the best agreement between the numerical solutions of the Fokker-Planck equation and the pdfs computed directly from the data.

In detail, in section \ref{fokker-planck-equation} the basic features for stochastic processes evolving in scale are discussed. A description of the optimisation routines is given in section \ref{sec_optimisation}. First results for turbulence data are shown in section \ref{sec_results_turbulence} and new insights are pointed out for this type of data. Further applications of the discussed methods are shown in section \ref{sec_applications}. We finish with some concluding remarks in section \ref{sec_conclusions}.

\section{\label{fokker-planck-equation}Fokker-Planck-equation}

We start with a situation, where for a complex system some amount of data $x(t)$ is given. Here $x$ denotes the describing quantity, such as heights for surfaces or velocity for turbulent fields and $t$ denotes a time or a space variable. For simplicity we assume that $x$ is a one dimensional quantity, noting that higher dimensional systems can be treated in a similar way \cite{siefert2006}. The scale dependent features are described by $y(t,\tau)$, where $\tau$ denotes the selected scale and $y$ a quantity describing the disorder (complexity) of $x(t)$ in a $\tau$-neighbourhood. $y$ may be a wavelet, a local roughness or any other local quantity (see for example \cite{karth2003}). Here we define $y$ as a simple increment
\begin{eqnarray}
\label{eq_increment}
y(t,\tau):=x(t+\tau)-x(t).
\end{eqnarray}
In order to obtain a statistically complete description of the system with respect to $y$, the joint probability density function $p(y_1,\tau_1,...,y_n,\tau_n)$ of $y(\tau)$ at different scales $\tau_i$, has to be known. The joint pdf is constructed from the set of $y_i(\tau_i)$ obtained at the same $t$ value. In the following the joint statistics of these increment processes are considered. Because of the involved scales the dimension of the joint probability density can be very high. Therefore it is in general very difficult to compute this joint pdf from empirical time series. However the description and computation can be highly simplified if Markov properties hold. This is the case if 
\begin{eqnarray}
\label{eq_markov_properites}
p(y_i,\tau_i|y_{i+1},\tau_{i+1},...,y_n,\tau_n) = p(y_i,\tau_i|y_{i+1},\tau_{i+1})
\end{eqnarray}
is true for all $i$ and $n > i$. Without loss of generality we take $\tau_{i} < \tau_{i+1}$. It should be noted that the Markov property can be tested for a given data set \cite{renner01b,renner01,friedrich98,lueck2006, marcq01}. For valid Markov properties the joint probability density can be substantially simplified:
\begin{eqnarray}
\label{eq_probability_factor}
\lefteqn{p(y_1,\tau_1,...,y_n,\tau_n) =  } \hspace{-0.0cm} \\
\nonumber \\
& & p(y_1,\tau_1,|y_2,\tau_2)\cdot ...\cdot p(y_{n-1},\tau_{n-1}|y_n,\tau_n)\cdot p(y_n,\tau_n) .\nonumber 
\end{eqnarray}
Because the conditional pdfs of first order (the right side of Eq.~(\ref{eq_markov_properites})) provide a complete description of a Markov process, they are the basic quantity to measure the correctness of the description of a Markov process. This issue and the importance of using conditional pdfs and not unconditional pdfs for the verification of the estimated process are discussed in \cite{renner01, renner02a, friedrich00b,renner01b}.

The idea proposed in \cite{friedrich97,friedrich97b,renner01b,renner01} is to model these conditional pdfs of first order with a Fokker-Planck equation \cite{risken96,gardiner85} evolving in scale,
\begin{eqnarray}
\label{eq_fokker_planck}
\lefteqn{- \tau \frac{\partial}{\partial \tau} p(y,\tau|y_0,\tau_0) =  } \hspace{-0.0cm} \\ \nonumber \\
& & \left [ - \frac{\partial}{\partial y}  D^{(1)}(y,\tau) + \frac{\partial^2}{\partial y^2}  D^{(2)}(y,\tau) \right ] p(y,\tau|y_0,\tau_0) .\nonumber 
\end{eqnarray}
Note that, in contrast to the usual definition of the Fokker-Planck equation, here both sides are multiplied by $\tau$. This corresponds to a logarithmic length scale as used in \cite{friedrich97}. This choice is convenient for analysing fractal scaling features, but does not imply any loss of generality. $D^{(1)}(x,\tau)$ and $D^{(2)}(x,\tau)$ are the drift or diffusion coefficients, respectively, and are defined as
\begin{eqnarray}
\label{eq_def_kmcoeff}
\lefteqn{ D^{(i)}(x,\tau) =  } \hspace{-0.0cm} \\ \nonumber \\
& & \lim_{\Delta \tau \rightarrow 0} \frac{\tau}{i! \Delta \tau} \int \limits_{-\infty}^{+\infty} (x' -x)^i p(x',\tau - \Delta \tau | x,\tau) dx' .\nonumber 
\end{eqnarray}
It should be noted that the conditional pdf \parbox{2.6cm}{$ p(x',\tau-\Delta \tau | x,\tau)$} can be estimated directly from the data, and therefore the Kramers-Moyal coefficients can be determined by using Eq.~(\ref{eq_def_kmcoeff}). To see the validity of the Fokker-Planck ansatz the size of the 4th order Kramers-Moyal coefficient can be estimated, for further details see \cite{renner01,risken96,tutkun2004}.

\section{\label{sec_optimisation}Optimisation}

For the optimised estimation of the Kramers-Moyal coefficients an implementation of the limited-memory Broyden-Fletcher-Goldfarb-Shanno algorithm for constraint problems (L-BFGS-B algorithm) \cite{nocedal999,byrd1995,zhu1997} in R \cite{r-project} is used, which is described in the appendix in detail. The starting point is the approximation of the Kramers-Moyal coefficients determined by the evaluation of Eq.~(\ref{eq_def_kmcoeff}). The coefficients are approximated by functions with free parameters $q_i^{(j)}$
\begin{eqnarray}
\label{eq_kramers_parametrization}
\tilde{D}^{(1)}(y,\tau) = f(y,\tau,q_0^{(1)},...,q_{\tilde{m}}^{(1)})\\
\tilde{D}^{(2)}(y,\tau) = g(y,\tau,q_0^{(2)},...,q_{\tilde{n}}^{(2)}).
\end{eqnarray}
Solving Eq.~(\ref{eq_fokker_planck}) as proposed in \cite{renner01} by using $\tilde{D}^{(1)}(y,\tau)$ and $\tilde{D}^{(2)}(y,\tau)$ as the drift and diffusion coefficient respectively, leads to a conditional pdf of first order $p_{num}(y_{i-1},\tau_{i-1} | y_i,\tau_i,q_0^{(1)},...,q_{\tilde{m}}^{(1)},q_0^{(2)},...,q_{\tilde{n}}^{(2)})$. In order to maximise the agreement between these numerical solutions $p_{num}$ of Eq.~(\ref{eq_fokker_planck}) and the pdfs from the empirical data, a measure is needed. Here, the weighted mean square error in logarithmic space is used, which is defined as 
\begin{eqnarray}
\label{eq_def_distance_MSE}
d_M(p_{n},p_{ref}) :=   \frac{\int \limits_{R}dr \; (p_{n}+p_{ref}) (\ln p_{n}-\ln p_{ref})^2}{\int \limits_{R}dr \; (p_{n}+p_{ref}) (\ln^2 p_{n}+\ln^2 p_{ref})}.
\end{eqnarray}
Here $R$ denotes the subspace, where an estimate of $p_{ref}$ from empirical data is possible and $p_{ref} > 0$. $p_n$ and $p_{ref}$ are joint probabilities of second order $p(y_{i-1},\tau_{i-1} ; y_i,\tau_i)$ which are obtained from the conditional probabilities by the multiplication with the empirical probability density $p(y_i,\tau_i)$. The L-BFGS-B algorithm minimizes the non-linear function $d_M(p_n,p_{ref},q)$ under constraints for each component of q, which may be denoted as $L \leq q \leq U$. $q$ is the $N_q = \tilde{m} + \tilde{n}$ dimensional vector of all the parameters $q_0^{(1)},...,q_{\tilde{m}}^{(1)},q_0^{(2)},...,q_{\tilde{n}}^{(2)}$ that are necessary to determine the functional form of the first two Kramers-Moyal coefficients for a given scale $\tau$. $L$ and $U$ represent the lower and upper bound on $q$, respectively. 

\section{\label{sec_results_turbulence}Results for turbulence}

The procedure described above, is now applied to experimental data. The data considered were obtained from a cryogenic axisymmetric helium gas jet at a Reynolds number of $7.6 \cdot 10^5$. The data set contain $1.6 \cdot 10^7$ measurements of the velocity in the center of a free jet, where the distance between the anemometer and the nozzle was 40D and the diameter D of the nozzle was 2 mm. For further details we refer to \cite{chanal2000}. This high Reynolds number data set have the benefit, that the region between the Markov-Einstein coherence length \cite{lueck2006} and the integral length spans a large interval of scales. This is important because below the Markov-Einstein coherence length Eq.~(\ref{eq_fokker_planck}) cannot be applied due to the missing Markov properties. Above the integral length the properties of the turbulent cascade are not present any more.

A first approximation of the Kramers-Moyal coefficients is determined by means of their definition in Eq.~(\ref{eq_def_kmcoeff}). Then these first estimates are used to reconstruct the conditional probability density. In order to assess the quality of the solution the distance $d_M$ between the probability density of the data on a certain scale $\tau$ and the reconstructed probability density by the estimated process equation is calculated. Thereby , the distance  $d_M(p_n,p_{ref})$ given in Eq.~(\ref{eq_def_distance_MSE}) is used. Now the iterative algorithm, described in the appendix is used to minimize $d_M(p_n,p_{ref})$ with respect to $q$.

The optimisation, for the turbulence data used here, is performed for each value of the scale individually. In order to do this, the range of scales is divided into small half-open intervals $[\tau_{is},\tau_{ie}[$. Thus we apply here a piecewise constant approximation to the scale dependent process. The optimisation is performed independently for each of these intervals, with a parametrisation of the Kramers-Moyal coefficients that is no longer dependent on the scale, as was found in our previous work \cite{renner01, renner02a}. The first estimate of the Kramers-Moyal coefficients can be fitted with a polynomial of first order for the drift coefficient and a polynomial of second order for the diffusion coefficient. Therefore the Kramers-Moyal coefficients are parameterised as
\begin{eqnarray}
\label{eq_def_parametrization}
\tilde{D}^{(1)}(y,\tau) & = & q_0^{(1)} + q_1^{(1)} y \\
\tilde{D}^{(2)}(y,\tau) & = & q_0^{(2)} + q_1^{(2)} y + q_2^{(2)} y^2.
\end{eqnarray}
For many applications such a parametrisation of the  Kramers-Moyal coefficient with polynomials seems to be a good choice.

Though working with constant coefficients $q_i^{(j)}$ constitutes an approximation it has two advantages. The first is that the smoothness of the resulting functions $q_i^{(j)}$ with respect to $\tau$ provides a first assessment of the robustness of the optimisation process, if we assume the true coefficients to be smooth functions with respect to the scale. The second and more important advantage is, that the number of variables $N_q$, or in other words, the dimension of the space where the optimisation takes place, is smaller. This is important, because the optimisation in a lower dimensional space can be much faster than one in a high dimensional space. In addition the number of local minima may increase rapidly with the addition of more variables and therefore the localization of the global minimum becomes more difficult. 

The optimisation is performed for scales ranging from $\tau_{Markov}$ to 80750 sample steps, where $\tau_{Markov}$ denotes the Markov-Einstein coherence length, which is in this case 8 sample steps. The integral length for this data set is 715 sample steps. The scale intervals are chosen here in such a way, that
\begin{eqnarray}
\label{eq_delta_tau}
\tau_{ie} = max \left\{ \frac{\tau_{is}}{0.9};\tau_{is}+\tau_{Markov}\right \},
\end{eqnarray}
where $\tau_{is}$ denotes the left and $\tau_{ie}$ the right border of the scale interval. The initial estimate of the Kramers-Moyal coefficients by means of their definition in Eq.~(\ref{eq_def_kmcoeff}) can only be performed for a scale larger than $\tau_c$, where $\tau_c \geq \tau_{Markov}$. In our case $\tau_c = 5 \cdot \tau_{Markov}$. This is due to the procedure to perform the limit in Eq.~(\ref{eq_def_kmcoeff}) numerically, for more details see e.g. \cite{renner01}. Therefore as initial estimates the values of the Kramers-Moyal coefficients at the scale $\tau_c$ are used for scales $\tau \leq \tau_c$. The limit in Eq.~(\ref{eq_def_kmcoeff}) is calculated without the use of possible refinements in order to test the robustness of the optimisation routine. The boundaries $L$ and $U$ are set in a simple way, to prevent $\tilde{D}^{(2)}(y,\tau)$ from becoming negative. The average number of iterations before the optimisation stopped was around 25.

\begin{figure}
\includegraphics[width= 8.5cm]{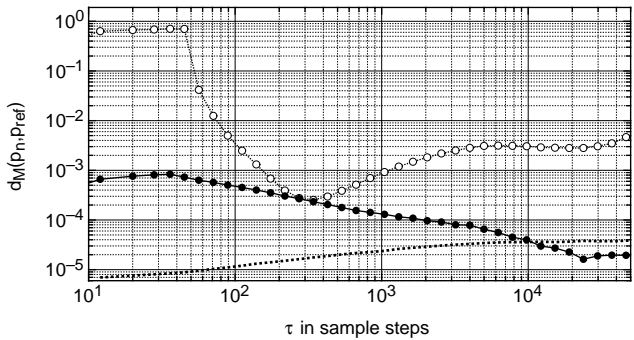}
\caption[$d_M$ between the pdfs of original and reconstructed data]{\label{fig_gm1h_125SU_distance}Distance measure $d_M$ between the pdfs of the original data and the reconstructed ones using the numerical solutions of the Fokker-Planck equation determined by the initial estimate of the Kramers-Moyal coefficients (open symbols) and optimised estimate of the Kramers-Moyal coefficients (black dots). The dotted line provides the expected distance, if both distributions have been produced by the same process.}
\end{figure}

The values of the distance measure between the pdfs of the original data and the reconstructed ones using the initial estimates of the Kramers-Moyal coefficients are displayed in Fig.~\ref{fig_gm1h_125SU_distance} as open symbols.
Three ranges can be identified. The first range spans from $\tau_{Markov}$ to $\tau_c$. Here the limit could not be calculated in a proper way and constant initial estimates of the Kramers-Moyal coefficients have been used, resulting in a nearly constant distance measure in this range. The second range spans from $\tau_c$ to $\tau_{op}$, where $\tau_{op}$ is around 300 sample steps. In this range the distance measure decreases monotonically with increasing scale $\tau$. This may be due to a better description of the data with increasing $\tau$, or due to a better performance of the initial estimate of the Kramers-Moyal coefficients, or due to both. In the third range the distance measure increases after the minimum at $\tau_{op}$, which marks the border between the second and the third range.

Performing the optimisation routine described above, the distance measure between the pdfs of the original data and the reconstructed ones using the optimisation routine is obtained. The distance measure for the optimised pdfs is displayed in Fig.~\ref{fig_gm1h_125SU_distance} as black dots. For very small scales $\tau \lesssim \tau_c$ the distance measure remains constant or increases slightly. For a very broad range of scales the distance measure then declines monotonically, until it saturates for very large scales. Interestingly a scale $\tau \sim \tau_{op}$ exists, where the distance function has approximately the same value for the initially estimated coefficients and the optimised one. This indicates that for $\tau_{op}$ the initial estimate of the Kramers-Moyal coefficients is already optimal. We obtained similar results for other data sets.

\begin{figure}
\includegraphics[width= 8.5cm]{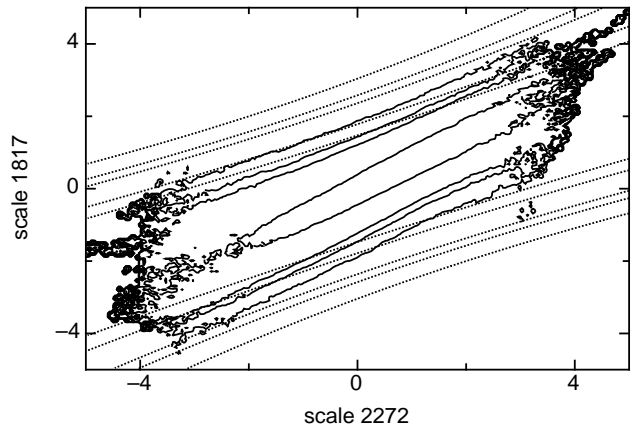}
\caption[Conditional pdfs of given data and numerical solutions] {\label{fig_condpdf_0}Conditional probability density $p(y(\tau = 1817)|y(\tau = 2272))$ of given data (unbroken lines) and reconstructed by the numerical solution of the Fokker-Planck equation (dotted lines) using the initial estimates of the Kramers-Moyal coefficients.}
\end{figure}

\begin{figure}
\includegraphics[width= 8.5cm]{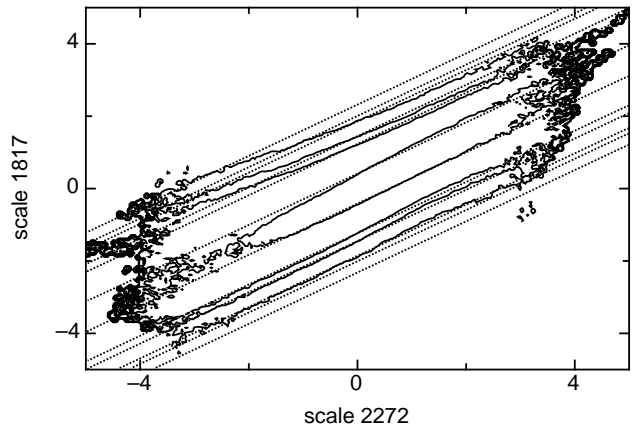}
\caption[onditional pdfs of given data and numerical solutions]{\label{fig_condpdf_final}Conditional probability density $p(y(\tau = 1817)|y(\tau = 2272))$ of given data (unbroken lines) and reconstructed by the numerical solution of the Fokker-Planck equation (dotted lines) using the optimised estimates of the Kramers-Moyal coefficients.}
\end{figure}

In order to assess the significance of the results the intrinsic error is estimated.  The data set is divided in sub sets and the distance between the distributions belonging to the corresponding sub sets is calculated. This is done for different sizes of sub sets and then extrapolated to obtain the intrinsic error for the whole data set. As seen in Fig.~\ref{fig_gm1h_125SU_distance} the intrinsic error is still smaller for scales up to $10^4$ than the distance measure for the optimised coefficients, nevertheless to our interpretation the magnitude of the distance measure is with $10^{-3}$ sufficiently small. see Fig.~\ref{fig_condpdf_0} and Fig.~\ref{fig_condpdf_final} for an example. 

\begin{figure}
\includegraphics[width= 8.5cm]{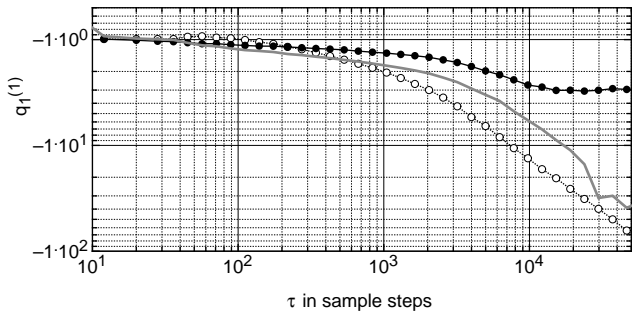}
\caption[Linear term of the drift coefficient]{\label{fig_d1_1}The parameter $q_1^{(1)}$. The initial estimate is denoted with white circles while the optimised one is denoted with black circles. The grey line shows the results for centered increments using optimised coefficents.}
\end{figure}

\begin{figure}
\includegraphics[width= 8.5cm]{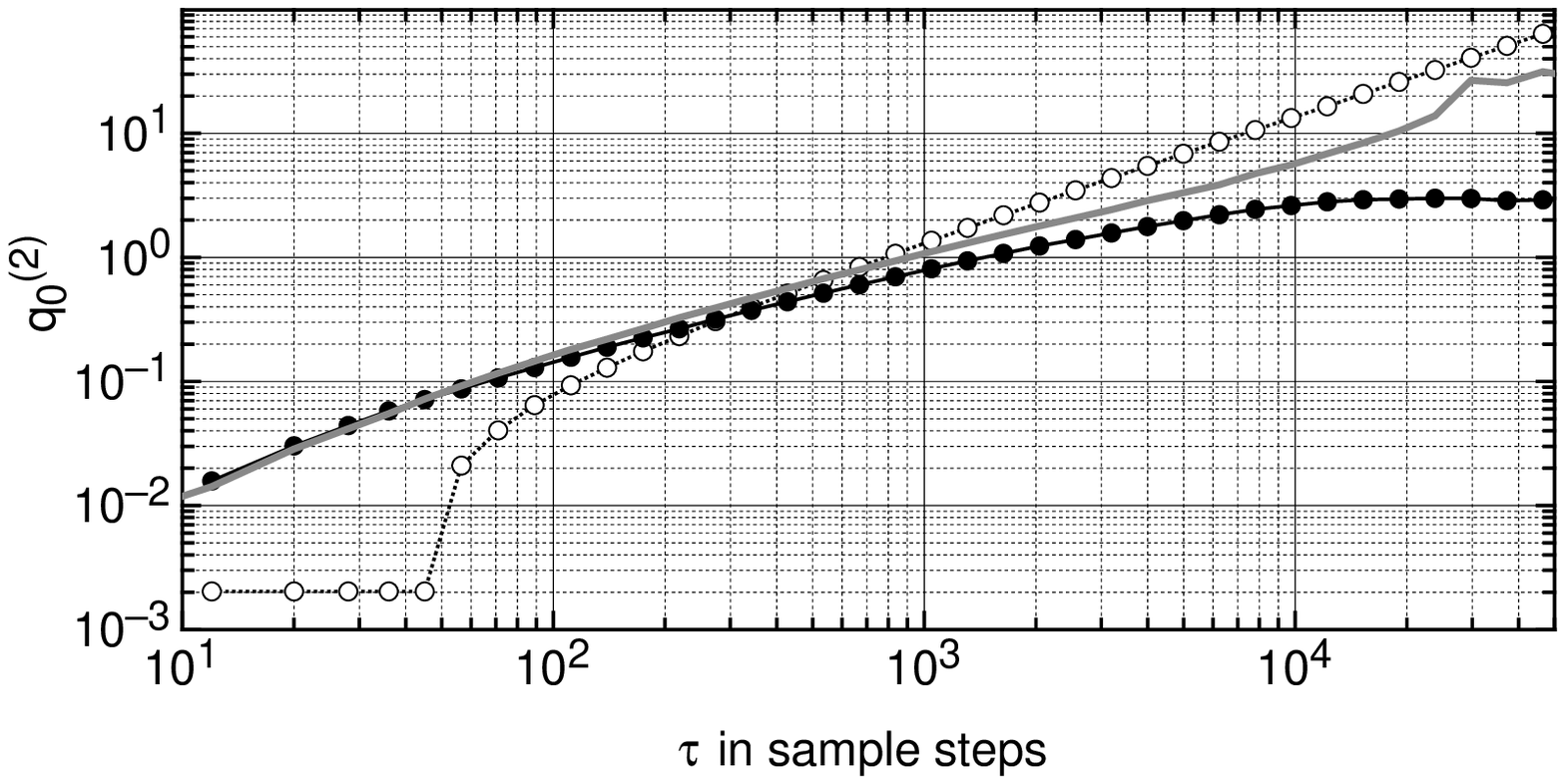}
\caption[Constant term of the diffusion coefficient]{\label{fig_d2_0}The parameter $q_0^{(2)}$. The initial estimate is denoted with white circles while the optimised one is denoted with black circles. The grey line shows the results for centered increments using optimised coefficents.}
\end{figure}

\begin{figure}
\includegraphics[width= 8.5cm]{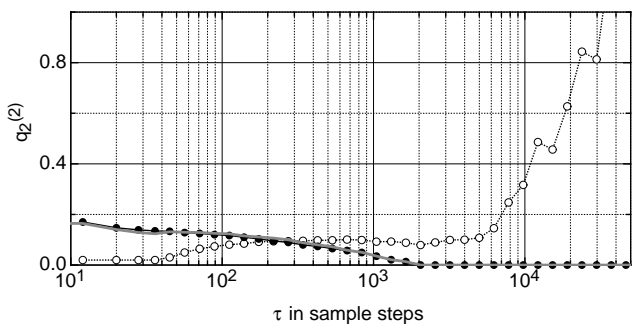}
\caption[Quadratic term of the diffusion coefficient]{\label{fig_d2_2}The parameter $q_2^{(2)}$. The initial estimate is denoted with white circles while the optimised one is denoted with black circles. The grey line shows the results for centered increments using optimised coefficents.}
\end{figure}

The graphs for the optimised coefficients $q_1^{(1)}$,  $q_0^{(2)}$ and $q_2^{(2)}$ are shown in Fig.~\ref{fig_d1_1}~-~\ref{fig_d2_2}.
For $q_0^{(1)}$ and $q_1^{(2)}$ the initial estimates as well as the optimised values are essentially equal to zero. This result has an interesting physical context. It has been shown that for a higher dimensional analysis a corresponding non-vanishing $q_1^{(2)}$ term violates the second von K\'arm\'an equation (see Eqs. (31) and (32) in \cite{siefert2006}).
In Fig.~\ref{fig_d2_2} two ranges can be identified. For smaller scales $\tau < \tau_s$, , where $\tau_s$ is around 2000 sample steps, $q_2^{(2)}$ takes non-vanishing positive values, while it becomes zero for scales $\tau > \tau_s$.  The other non-vanishing term of the second Kramers-Moyal coefficient, $q_0^{(2)}$, in Fig.~\ref{fig_d2_0}, exhibits in the same region a power-law behaviour which saturates for larger scales. The same is true for $q_1^{(1)}$ in Fig.~\ref{fig_d1_1}, but with a much higher accuracy. The exponent for $q_1^{(1)}$ in this region is $0.069$.

\begin{figure}
\includegraphics[width= 8.5cm]{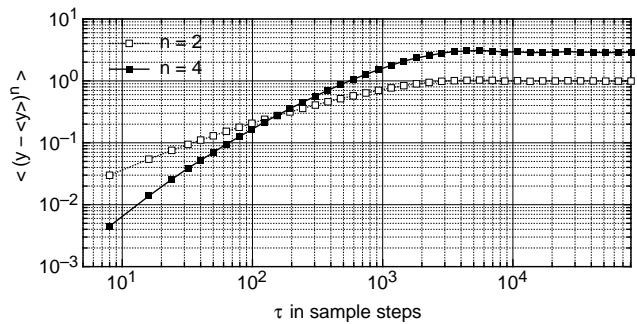}
\caption[Second and fourth moment of the data set]{\label{fig_moments}The second and fourth moment of the data set.}
\end{figure}

For the interpretation of $\tau_s$ we note, that $q_2^{(2)}$ represents the multiplicative noise in the system. By investigating the moments of the system, which are also called structure functions,  in Fig.~\ref{fig_moments}, it can be noted that the moments start to saturate at a scale, which is comparable to $\tau_s$. This indicates that $\tau_s$ is related to the integral length of the system. Further $q_2^{(2)}=0$ for $\tau > \tau_s$ is in agreement with  a Gaussian shape of the pdf of velocity increments for large scales.

\section{\label{sec_applications}Applications}

\begin{figure}
\includegraphics[width= 8.5cm]{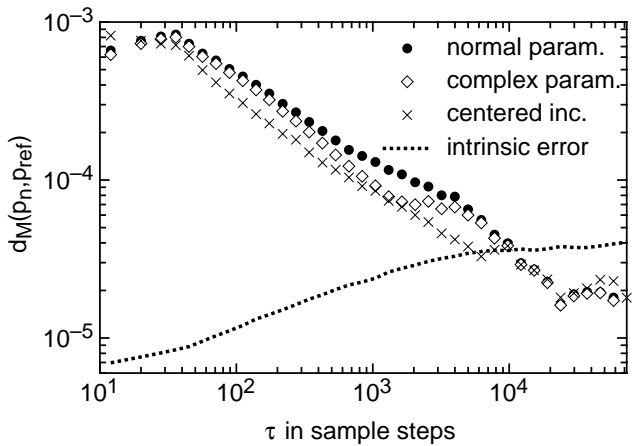}
\caption[Comparison of the distance function $d_M$ for different settings]{\label{fig_error_comparison}Comparison of the distance function $d_M$ for different settings. The dottetd line provides the expected distance, if both distributions have been produced by the same process.}
\end{figure}

The method above provides a much better answer to the central question of determining the correct Kramers-Moyal coefficients. But besides this it enables us to discuss further important questions arising from the description of scale dependent systems with a Fokker-Planck equation. The first of these questions is the optimal increment definition for the stochastic process as given by Eq.~(\ref{eq_increment}). We started our analysis using the left-justified increments, which are more common in the literature. Using left-justified increments means, that the smaller increment is nested inside the larger increment and that both increments have the left endpoint in common. For certain classes of systems this may introduce additional correlations between the increments that are not desired \cite{waechter2004b}. Thus it has been proposed to use centered increments instead of left-justified ones
\begin{eqnarray}
\label{eq_increment2}
y(t,\tau):=x\left(t+\frac{\tau}{2}\right)-x\left(t-\frac{\tau}{2}\right).
\end{eqnarray}
It is now possible to investigate the improvement of the description of the system by using centered increments. As a criterion we use the distance between the numerical solution of the Fokker-Planck equation, where the coefficients have been optimised, and the empirical pdf. As can be seen in Fig.~\ref{fig_error_comparison}, the distance measure exhibts smaller values for centered increments with the exceptions of very small and very large scales. This indicates that the description of this special system indeed can be improved using centered increments.
Reanalysing the coefficients that are shown in Fig.~\ref{fig_d1_1}~-~\ref{fig_d2_2} no principal changes are found for the use of centered increments. In Fig.~\ref{fig_d1_1}~-~\ref{fig_d2_2} additionally the results for the centered increment analysis are shown. The biggest change is found for $q_1^{(1)}$ and $q_0^{(2)}$ and for scales larger than the integral scale, which are less important.

The second important question concerns the parametrisation of the Kramers-Moyal coefficients. It is now possible to determine, whether a more complex parametrisation of the Kramers-Moyal coefficients, for example by using higher order polynomials, yields a better description of the system. As a simple example to illustrate this, the question of asymmetric Kramers-Moyal coefficients is considered. If the initial estimates of the Kramers-Moyal coefficients are examined, the functional form appears to be asymmetric in some cases. This seems to be especially true for very large scales, where the number of independent events becomes smaller. In order to verify if the underlying stochastic process can be better described by a separate parametrisation for negative and positive increments, the following parametrisation is chosen for the optimisation:
\begin{eqnarray}
\label{eq_def_parametrization2}
\tilde{D}^{(1)}(y,\tau) & = & \left \{ \begin{array}{lll}
	q_0^{(1)} + q_1^{(1)} y  	&  \textrm{if $y < 0$} \\
	q_2^{(1)} + q_3^{(1)} y  	&  \textrm{if $y \geq 0$}
\end{array} \right.\\
\tilde{D}^{(2)}(y,\tau) & = & \left \{ \begin{array}{lll}
	q_0^{(2)} + q_1^{(2)} y + q_2^{(2)} y^2  	&  \textrm{if $y < 0$} \\
	q_3^{(2)} + q_4^{(2)} y + q_5^{(2)} y^2  	&  \textrm{if $y \geq 0$}
\end{array} \right.
\end{eqnarray}
As depicted in Fig.~\ref{fig_error_comparison} the distance function takes smaller values than for the original optimisation, although the improvement is not as large as when using centered increments. It should further be noted, that an improvement is in this case not surprising since the optimisation is now performed in a higher dimensional space and the space used for the original optimisation is a subspace of this second optimisation. Nevertheless this finding is in accordance with the proposed importance of higher odd order terms in the diffusion coefficient \cite{marcq01}.

Therefore it may be inferred that a further improvement of the description of this system may be provided by using an appropriate increment definition rather than adopting the assumption of asymmetric Kramers-Moyal coefficients. The first provides a better description of the system by using fewer free variables compared to the second. Another questions that may be answered in such a fashion is the use of higher order Kramers-Moyal coefficients, especially the fourth order coefficient because of its importance for the application of the theorem of Pawula \cite{risken96}.

\section{\label{sec_conclusions}Conclusions}

In this work we have shown a practical way to implement an optimisation routine to improve the description of hierarchical systems by means of a Fokker-Planck equation. In order to do so, first an estimate of the Kramers-Moyal coefficients using their definition in Eq.~(\ref{eq_def_kmcoeff}) is calculated. This initial estimate is then used to solve the Fokker-Planck equation numerically and to obtain as a solution the conditional probability density functions (pdfs) of first order. As a next step the distance between this reconstructed conditional probability and the one obtained directly from the time series is determined using  Eq.~(\ref{eq_def_distance_MSE}). This procedure forms the basis of our optimisation routine. A parametrisation of the initial estimate of the Kramers-Moyal coefficients is chosen, with a specified number of variables $N_q$. The L-BFGS-B algorithm is employed to minimize the distance between the numerical solutions of the Fokker-Planck equation and the empirical pdfs by adjusting the free variables. The L-BFGS-B algorithm is an algorithm which is very effective in the case of an optimisation of many variables which may be constrained. Therefore the method proposed here will also be effective for very complex parametrisation, as long as these parametrisations are not misspecified.

We applied the optimisation routine to a time series of velocity measurements obtained from a cryogenic axisymmetric helium gas jet. We demonstrated the benefits of this optimisation routine. At first it is possible to obtain values of the Kramers-Moyal coefficients for much smaller scales, due to the fact that it is no longer necessary to calculate a limit in scale which is the bottle-neck of the original Kramers-Moyal method. At second the optimised coefficients produce numerical solutions of the Fokker-Planck equation that are much closer to the empirical pdfs than those produced by the initial estimates. At third possible systematic errors in the classical estimation routine of the Kramers-Moyal coefficients that have been pointed out in the literature can be avoided using this optimisation routine. At fourth the optimised coefficients show remarkable simple functional forms in a large scaling region, while the behaviour of the initial estimates is much more ambiguous. At last the results produced by this optimisation routine are remarkable stable. Independent optimisations have been performed for small intervals in scale bordering on each other, producing estimates which are very smooth with respect to the scale. Therefore this method provides the means to determine the Kramers-Moyal coefficients with much more accuracy or to determine correct Kramers-Moyal coefficients for small data sets.

Possible applications for this refined approach have been shown. First the question of the appropriate increment definition has been considered. It has been shown that by using centered increments instead of left-justified ones, the description of the underlying stochastic process for our example system can be improved. Second the question of the optimal parametrisation of Kramers-Moyal coefficients in the Fokker-Planck equation has been considered. It was shown that in our case an asymmetric parametrisation provides only a slight improvement. This aspect interests because it is directly related to closure of the higher order moments, see Eq.~(4.13) in \cite{renner01}. With our findings here we see that a perturbing linear term for the diffusion coefficient may have no significance; thus the reported contradiction of the reconstructed Fokker-Planck equation with the second K\'arm\'an equation seems to have no significance, or saying it in other words, this discrepancy is just a result of an inaccurate estimation. Further applications may include the analysis of more complex parametrisations of the Kramers-Moyal coefficients and the influence of higher order Kramers-Moyal coefficients, thereby offering new insights in the complexity of turbulence.

\begin{acknowledgments}
We thank B. Chabaud and O. Chanal for providing us with excellent data. 
\end{acknowledgments}

\appendix*

\section{\label{ap_L-BFGS-B algorithm}L-BFGS-B algorithm}

For the optimised estimation of the Kramers-Moyal coefficients we apply an iterative procedure, called L-BFGS-B algorithm \cite{nocedal999,byrd1995,zhu1997}. Solving Eq.~(\ref{eq_fokker_planck}) as proposed in \cite{renner01} leads to a conditional pdf of first order $p_{num}(y_{i-1},\tau_{i-1} | y_i,\tau_i,q)$. In order to maximise the agreement between the numerical solutions $p_n$ and the pdfs from the empirical data $p_{ref}$, a measure is needed. Here $d_M(p_n,p_{ref},q)$ is used, which is defined in Eq.~(\ref{eq_def_distance_MSE}). The L-BFGS-B algorithm minimizes the non-linear function $d_M(p_n,p_{ref},q)$, here $d(q)$ is used as a short hand notation, under the constraint $L \leq q \leq U$. $L$ and $U$ represent the lower and upper bound on $q$, respectively.

Using the details provided above, an iterative procedure is started to find the vector $q$ which minimizes the distance function $d(q)$. As a first step of each iteration $(k)$ a quadratic model $m_{(k)}(q)$ of $d(q)$ at the iterate $q_{(k)}$ is computed,
\begin{eqnarray}
\label{eq_mk}
\lefteqn{m_{(k)}(q) :=   } \hspace{-0.0cm} \\ \nonumber \\
& & d(q_{(k)}) + g_{(k)}^T  (q-q_{(k)}) + \frac{1}{2}(q-q_{(k)})^T  B_{(k)}  (q-q_{(k)}), \nonumber 
\end{eqnarray}
where $g_{(k)}$ denotes the gradient of $d(q)$ with respect to $q$ and $B_{(k)}$ is a limited-memory BFGS approximation to the Hessian. 

Because the following steps have to be repeated for each ${(k)}$, the index ${(k)}$ is omitted for these steps if it does not change the meaning. As a second step a set of active bounds has to be found using the gradient projection method. The projection of an arbitrary point $q$ onto the feasible region is defined by
\begin{eqnarray}
\label{eq_def_projection}
P(q,L,U)_i = \left \{ \begin{array}{lll}
	L_i         &                       & \textrm{$q_i < L_i$,} \\
	q_i & \textrm{$\,$ if $\,$} & \textrm{$q_i \in [L_i,U_i]$,} \\
	U_i         &                       & \textrm{$q_i > U_i$.}
\end{array} \right.
\end{eqnarray}
Therefore a piecewise linear path $q(s)$, which is the projection of the steepest descent direction at the starting point $q^0$ onto the feasible region, defined by Eq.~(\ref{eq_def_projection}), is denoted by
\begin{eqnarray}
\label{eq_lin_path}
q(s) = P(q^0-sg,L,U).
\end{eqnarray}
As a third step the generalized Cauchy point $q^c$, which is defined as the first local minimizer of the function $m_{(k)}(q(s))$ on the piecewise linear path $q(s)$, is computed. The components of $q^c$ which are at their upper or lower bound, $U$ or $L$, comprise the active set $\mathcal{A}(q^c)$ of variables.

As a fourth step the following quadratic problem over the subspace of free variables is considered.
\begin{eqnarray}
\label{eq_subspace1}
\min \{ m_{(k)}(q) | q_i = q_i^c \;\; \forall \; i \in \mathcal{A}(q^c) \},
\end{eqnarray}
\begin{eqnarray}
\label{eq_subspace2}
\textrm{subject to} \; L \leq q \leq U \;\; \forall \; i \notin \mathcal{A}(q^c).
\end{eqnarray}
Eq.~(\ref{eq_subspace1}) is solved approximately without the condition of Eq.~(\ref{eq_subspace2}) using a direct primal method \cite{byrd1995}. The solution of this unconstrained problem is denoted by $\delta^u$. Therefore the solution of the constrained problem can be written as
\begin{eqnarray}
\label{eq_constrained_solution}
q_{(k+1),i}^s = \left \{ \begin{array}{lll}
	q_{(k),i}^c                     &  \textrm{if $i \notin \mathcal{F}$,} \\
	q_{(k),i}^c + (Z \delta^*)_{(k),i}  &  \textrm{if $i \in \mathcal{F}$.}
\end{array} \right.
\end{eqnarray}
where 
\begin{eqnarray}
\label{eq_def_alpha}
\delta^* = \alpha^* \delta^u
\end{eqnarray}
and
\begin{eqnarray}
\label{eq_calc_alpha}
\lefteqn{\alpha^* :=   } \hspace{-0.0cm} \\ \nonumber \\
& & \max\{\alpha | \alpha \leq 1, L_i - q_i^c \leq \alpha \delta_i^u \leq U_i - q_i^c, \; i \in \mathcal{F}\}. \nonumber
\end{eqnarray}
$Z$ denotes the $N_q \times N_q^c$ matrix of unit vectors (i.e. each column is a column of the identity matrix), that span the subspace of free variables at $q^c$, where $N_q^c$ denotes the number of free variables at $q^c$. $\mathcal{F}$ denotes the set of indices korresponding to the set of free variables.

As a last step a line search between the current $q_{(k)}$ and the approximate minimizer $q_{(k+1)}^s$ is performed, which satisfies the strong Wolfe conditions
\begin{eqnarray}
\label{eq_bfgs_strong_wolfe_cond1}
d(q_{k+1}) \leq d(q_{k}) + c_1 \lambda_k g_k^T (q_{k+1}^s - q_k)
\end{eqnarray}
\begin{eqnarray}
\label{eq_bfgs_strong_wolfe_cond2}
|g_{k+1}^T (q_{k+1}^s - q_k)| \leq c_2 |g_{k}^T (q_{k+1}^s - q_k)|
\end{eqnarray}
where $\lambda_{(k)}$ is the step length and $c_1=10^{-4}$ and $c_2=0.9$. Here the algorithm of More and Thuente \cite{more1994} is used. The solution of this line search is used as the next iterate $q_{(k+1)}$.

The optimisation procedure is stopped, if the value for $d(q)$ cannot be reduced by a certain percentage, which will be in our case around $10^{-8}$ or a certain number of iteration will be reached, which is 50.

\newpage 
\bibliography{AG_LITERATUR}

\begin{thebibliography}{38}
\expandafter\ifx\csname natexlab\endcsname\relax\def\natexlab#1{#1}\fi
\expandafter\ifx\csname bibnamefont\endcsname\relax
  \def\bibnamefont#1{#1}\fi
\expandafter\ifx\csname bibfnamefont\endcsname\relax
  \def\bibfnamefont#1{#1}\fi
\expandafter\ifx\csname citenamefont\endcsname\relax
  \def\citenamefont#1{#1}\fi
\expandafter\ifx\csname url\endcsname\relax
  \def\url#1{\texttt{#1}}\fi
\expandafter\ifx\csname urlprefix\endcsname\relax\def\urlprefix{URL }\fi
\providecommand{\bibinfo}[2]{#2}
\providecommand{\eprint}[2][]{\url{#2}}

\bibitem[{\citenamefont{Peinke et~al.}(2002)\citenamefont{Peinke, Renner, and
  Friedrich}}]{peinke2002}
\bibinfo{author}{\bibfnamefont{J.}~\bibnamefont{Peinke}},
  \bibinfo{author}{\bibfnamefont{C.}~\bibnamefont{Renner}}, \bibnamefont{and}
  \bibinfo{author}{\bibfnamefont{R.}~\bibnamefont{Friedrich}}, in
  \emph{\bibinfo{booktitle}{Complexity from Microscopic to Macroscopic Scales:
  Coherence and Large Deviations edited by A.T. Skjeltorp and T. Vicsek}}
  (\bibinfo{publisher}{NATO Science Series II Vol.63}, \bibinfo{year}{2002}),
  pp. \bibinfo{pages}{151--169}.

\bibitem[{\citenamefont{Siegert et~al.}(1998)\citenamefont{Siegert, Friedrich,
  and Peinke}}]{siegert98}
\bibinfo{author}{\bibfnamefont{S.}~\bibnamefont{Siegert}},
  \bibinfo{author}{\bibfnamefont{R.}~\bibnamefont{Friedrich}},
  \bibnamefont{and} \bibinfo{author}{\bibfnamefont{J.}~\bibnamefont{Peinke}},
  \bibinfo{journal}{Phys. Lett. A} \textbf{\bibinfo{volume}{243}},
  \bibinfo{pages}{275} (\bibinfo{year}{1998}).

\bibitem[{\citenamefont{Friedrich et~al.}(1998)\citenamefont{Friedrich, Zeller,
  and Peinke}}]{friedrich98}
\bibinfo{author}{\bibfnamefont{R.}~\bibnamefont{Friedrich}},
  \bibinfo{author}{\bibfnamefont{J.}~\bibnamefont{Zeller}}, \bibnamefont{and}
  \bibinfo{author}{\bibfnamefont{J.}~\bibnamefont{Peinke}},
  \bibinfo{journal}{Europhys. Lett.} \textbf{\bibinfo{volume}{41}},
  \bibinfo{pages}{153} (\bibinfo{year}{1998}).

\bibitem[{\citenamefont{Friedrich
  et~al.}(2000{\natexlab{a}})\citenamefont{Friedrich, Siegert, Peinke, Lueck,
  Siefert, Lindemann, Raethjen, Deuschl, and Pfister}}]{friedrich00}
\bibinfo{author}{\bibfnamefont{R.}~\bibnamefont{Friedrich}},
  \bibinfo{author}{\bibfnamefont{S.}~\bibnamefont{Siegert}},
  \bibinfo{author}{\bibfnamefont{J.}~\bibnamefont{Peinke}},
  \bibinfo{author}{\bibfnamefont{S.}~\bibnamefont{Lueck}},
  \bibinfo{author}{\bibfnamefont{M.}~\bibnamefont{Siefert}},
  \bibinfo{author}{\bibfnamefont{M.}~\bibnamefont{Lindemann}},
  \bibinfo{author}{\bibfnamefont{J.}~\bibnamefont{Raethjen}},
  \bibinfo{author}{\bibfnamefont{G.}~\bibnamefont{Deuschl}}, \bibnamefont{and}
  \bibinfo{author}{\bibfnamefont{G.}~\bibnamefont{Pfister}},
  \bibinfo{journal}{Phys. Lett. A} \textbf{\bibinfo{volume}{271}},
  \bibinfo{pages}{217} (\bibinfo{year}{2000}{\natexlab{a}}).

\bibitem[{\citenamefont{Siefert et~al.}(2003)\citenamefont{Siefert, Kittel,
  Friedrich, and Peinke}}]{siefert03}
\bibinfo{author}{\bibfnamefont{M.}~\bibnamefont{Siefert}},
  \bibinfo{author}{\bibfnamefont{A.}~\bibnamefont{Kittel}},
  \bibinfo{author}{\bibfnamefont{R.}~\bibnamefont{Friedrich}},
  \bibnamefont{and} \bibinfo{author}{\bibfnamefont{J.}~\bibnamefont{Peinke}},
  \bibinfo{journal}{Europhys. Lett.} \textbf{\bibinfo{volume}{61}},
  \bibinfo{pages}{466} (\bibinfo{year}{2003}).

\bibitem[{\citenamefont{Waechter et~al.}(2004)\citenamefont{Waechter,
  Kouzmitchev, and Peinke}}]{waechter2004b}
\bibinfo{author}{\bibfnamefont{M.}~\bibnamefont{Waechter}},
  \bibinfo{author}{\bibfnamefont{A.}~\bibnamefont{Kouzmitchev}},
  \bibnamefont{and} \bibinfo{author}{\bibfnamefont{J.}~\bibnamefont{Peinke}},
  \bibinfo{journal}{Physical Review E} \textbf{\bibinfo{volume}{70}}
  (\bibinfo{year}{2004}).

\bibitem[{\citenamefont{Frank et~al.}(2004)\citenamefont{Frank, Beek, and
  Friedrich}}]{frank2004}
\bibinfo{author}{\bibfnamefont{T.~D.} \bibnamefont{Frank}},
  \bibinfo{author}{\bibfnamefont{P.~J.} \bibnamefont{Beek}}, \bibnamefont{and}
  \bibinfo{author}{\bibfnamefont{R.}~\bibnamefont{Friedrich}},
  \bibinfo{journal}{Physics Letters A} \textbf{\bibinfo{volume}{328}},
  \bibinfo{pages}{219} (\bibinfo{year}{2004}).

\bibitem[{\citenamefont{Kriso et~al.}(2002)\citenamefont{Kriso, Peinke,
  Friedrich, and Wagner}}]{kriso02}
\bibinfo{author}{\bibfnamefont{S.}~\bibnamefont{Kriso}},
  \bibinfo{author}{\bibfnamefont{J.}~\bibnamefont{Peinke}},
  \bibinfo{author}{\bibfnamefont{R.}~\bibnamefont{Friedrich}},
  \bibnamefont{and} \bibinfo{author}{\bibfnamefont{P.}~\bibnamefont{Wagner}},
  \bibinfo{journal}{Phys. Lett. A} \textbf{\bibinfo{volume}{299}},
  \bibinfo{pages}{287} (\bibinfo{year}{2002}).

\bibitem[{\citenamefont{Kuusela}(2004)}]{kuusela04}
\bibinfo{author}{\bibfnamefont{T.}~\bibnamefont{Kuusela}},
  \bibinfo{journal}{Phys. Rev. E} \textbf{\bibinfo{volume}{69}},
  \bibinfo{pages}{031916} (\bibinfo{year}{2004}).

\bibitem[{\citenamefont{Ghasemi et~al.}(2006)\citenamefont{Ghasemi, Sahimi,
  Peinke, and Tabar}}]{ghasemi2006}
\bibinfo{author}{\bibfnamefont{F.}~\bibnamefont{Ghasemi}},
  \bibinfo{author}{\bibfnamefont{M.}~\bibnamefont{Sahimi}},
  \bibinfo{author}{\bibfnamefont{J.}~\bibnamefont{Peinke}}, \bibnamefont{and}
  \bibinfo{author}{\bibfnamefont{M.~R.~R.} \bibnamefont{Tabar}},
  \bibinfo{journal}{Journal of Biological Physics}
  \textbf{\bibinfo{volume}{32}}, \bibinfo{pages}{117} (\bibinfo{year}{2006}).

\bibitem[{\citenamefont{Waechter et~al.}(2003)\citenamefont{Waechter, Riess,
  Kantz, and Peinke}}]{waechter03}
\bibinfo{author}{\bibfnamefont{M.}~\bibnamefont{Waechter}},
  \bibinfo{author}{\bibfnamefont{F.}~\bibnamefont{Riess}},
  \bibinfo{author}{\bibfnamefont{H.}~\bibnamefont{Kantz}}, \bibnamefont{and}
  \bibinfo{author}{\bibfnamefont{J.}~\bibnamefont{Peinke}},
  \bibinfo{journal}{Europhysics Letters} pp. \bibinfo{pages}{579--585}
  (\bibinfo{year}{2003}).

\bibitem[{\citenamefont{Jafari et~al.}(2003)\citenamefont{Jafari, Fazeli,
  Ghasemi, Allaei, Tabar, zad, and Kavei}}]{jafari2003}
\bibinfo{author}{\bibfnamefont{G.}~\bibnamefont{Jafari}},
  \bibinfo{author}{\bibfnamefont{S.}~\bibnamefont{Fazeli}},
  \bibinfo{author}{\bibfnamefont{F.}~\bibnamefont{Ghasemi}},
  \bibinfo{author}{\bibfnamefont{S.~V.} \bibnamefont{Allaei}},
  \bibinfo{author}{\bibfnamefont{M.~R.~R.} \bibnamefont{Tabar}},
  \bibinfo{author}{\bibfnamefont{A.~I.} \bibnamefont{zad}}, \bibnamefont{and}
  \bibinfo{author}{\bibfnamefont{G.}~\bibnamefont{Kavei}},
  \bibinfo{journal}{Physical Review Letters} \textbf{\bibinfo{volume}{91}}
  (\bibinfo{year}{2003}).

\bibitem[{\citenamefont{Friedrich and
  Peinke}(1997{\natexlab{a}})}]{friedrich97b}
\bibinfo{author}{\bibfnamefont{R.}~\bibnamefont{Friedrich}} \bibnamefont{and}
  \bibinfo{author}{\bibfnamefont{J.}~\bibnamefont{Peinke}},
  \bibinfo{journal}{Phys. Rev. Lett.} \textbf{\bibinfo{volume}{78}},
  \bibinfo{pages}{863} (\bibinfo{year}{1997}{\natexlab{a}}).

\bibitem[{\citenamefont{Renner et~al.}(2001{\natexlab{a}})\citenamefont{Renner,
  Peinke, and Friedrich}}]{renner01}
\bibinfo{author}{\bibfnamefont{C.}~\bibnamefont{Renner}},
  \bibinfo{author}{\bibfnamefont{J.}~\bibnamefont{Peinke}}, \bibnamefont{and}
  \bibinfo{author}{\bibfnamefont{R.}~\bibnamefont{Friedrich}},
  \bibinfo{journal}{J. Fluid Mech.} \textbf{\bibinfo{volume}{433}},
  \bibinfo{pages}{383} (\bibinfo{year}{2001}{\natexlab{a}}).

\bibitem[{\citenamefont{Tutkun and Mydlarski}(2006)}]{tutkun2004}
\bibinfo{author}{\bibfnamefont{M.}~\bibnamefont{Tutkun}} \bibnamefont{and}
  \bibinfo{author}{\bibfnamefont{L.}~\bibnamefont{Mydlarski}},
  \bibinfo{journal}{New Journal of Physics} \textbf{\bibinfo{volume}{6}}
  (\bibinfo{year}{2006}).

\bibitem[{\citenamefont{Friedrich
  et~al.}(2000{\natexlab{b}})\citenamefont{Friedrich, Peinke, and
  Renner}}]{friedrich00b}
\bibinfo{author}{\bibfnamefont{R.}~\bibnamefont{Friedrich}},
  \bibinfo{author}{\bibfnamefont{J.}~\bibnamefont{Peinke}}, \bibnamefont{and}
  \bibinfo{author}{\bibfnamefont{C.}~\bibnamefont{Renner}},
  \bibinfo{journal}{Phys. Rev. Lett.} \textbf{\bibinfo{volume}{84}},
  \bibinfo{pages}{5224} (\bibinfo{year}{2000}{\natexlab{b}}).

\bibitem[{\citenamefont{Renner et~al.}(2001{\natexlab{b}})\citenamefont{Renner,
  Peinke, and Friedrich}}]{renner01b}
\bibinfo{author}{\bibfnamefont{C.}~\bibnamefont{Renner}},
  \bibinfo{author}{\bibfnamefont{J.}~\bibnamefont{Peinke}}, \bibnamefont{and}
  \bibinfo{author}{\bibfnamefont{R.}~\bibnamefont{Friedrich}},
  \bibinfo{journal}{Physica A} \textbf{\bibinfo{volume}{298}},
  \bibinfo{pages}{499} (\bibinfo{year}{2001}{\natexlab{b}}).

\bibitem[{\citenamefont{M.Ausloos and Ivanova}(2003)}]{ausloos2003}
\bibinfo{author}{\bibnamefont{M.Ausloos}} \bibnamefont{and}
  \bibinfo{author}{\bibfnamefont{K.}~\bibnamefont{Ivanova}},
  \bibinfo{journal}{Physical Review E} \textbf{\bibinfo{volume}{68}}
  (\bibinfo{year}{2003}).

\bibitem[{\citenamefont{Nawroth and Peinke}(2006)}]{nawroth2006a}
\bibinfo{author}{\bibfnamefont{A.~P.} \bibnamefont{Nawroth}} \bibnamefont{and}
  \bibinfo{author}{\bibfnamefont{J.}~\bibnamefont{Peinke}},
  \bibinfo{journal}{Physics Letters A} \textbf{\bibinfo{volume}{360}},
  \bibinfo{pages}{234} (\bibinfo{year}{2006}).

\bibitem[{\citenamefont{Ragwitz and Kantz}(2001)}]{ragwitz01}
\bibinfo{author}{\bibfnamefont{M.}~\bibnamefont{Ragwitz}} \bibnamefont{and}
  \bibinfo{author}{\bibfnamefont{H.}~\bibnamefont{Kantz}},
  \bibinfo{journal}{Phys. Rev. Lett.} \textbf{\bibinfo{volume}{87}}
  (\bibinfo{year}{2001}).

\bibitem[{\citenamefont{Friedrich et~al.}(2002)\citenamefont{Friedrich, Renner,
  Siefert, and Peinke}}]{friedrich02}
\bibinfo{author}{\bibfnamefont{R.}~\bibnamefont{Friedrich}},
  \bibinfo{author}{\bibfnamefont{C.}~\bibnamefont{Renner}},
  \bibinfo{author}{\bibfnamefont{M.}~\bibnamefont{Siefert}}, \bibnamefont{and}
  \bibinfo{author}{\bibfnamefont{J.}~\bibnamefont{Peinke}},
  \bibinfo{journal}{Phys. Rev. Lett.} \textbf{\bibinfo{volume}{89}},
  \bibinfo{pages}{149401} (\bibinfo{year}{2002}).

\bibitem[{\citenamefont{Ragwitz and Kantz}(2002)}]{ragwitz02b}
\bibinfo{author}{\bibfnamefont{M.}~\bibnamefont{Ragwitz}} \bibnamefont{and}
  \bibinfo{author}{\bibfnamefont{H.}~\bibnamefont{Kantz}},
  \bibinfo{journal}{Phys. Rev. Lett.} \textbf{\bibinfo{volume}{89}},
  \bibinfo{pages}{149402} (\bibinfo{year}{2002}).

\bibitem[{\citenamefont{Sura and Barsugli}(2002)}]{sura02}
\bibinfo{author}{\bibfnamefont{P.}~\bibnamefont{Sura}} \bibnamefont{and}
  \bibinfo{author}{\bibfnamefont{J.}~\bibnamefont{Barsugli}},
  \bibinfo{journal}{Phys. Lett. A} \textbf{\bibinfo{volume}{305}},
  \bibinfo{pages}{304} (\bibinfo{year}{2002}).

\bibitem[{\citenamefont{Kleinhans et~al.}(2005)\citenamefont{Kleinhans,
  Friedrich, Nawroth, and Peinke}}]{kleinhans2005}
\bibinfo{author}{\bibfnamefont{D.}~\bibnamefont{Kleinhans}},
  \bibinfo{author}{\bibfnamefont{R.}~\bibnamefont{Friedrich}},
  \bibinfo{author}{\bibfnamefont{A.}~\bibnamefont{Nawroth}}, \bibnamefont{and}
  \bibinfo{author}{\bibfnamefont{J.}~\bibnamefont{Peinke}},
  \bibinfo{journal}{Physics Letters A} \textbf{\bibinfo{volume}{346}},
  \bibinfo{pages}{42} (\bibinfo{year}{2005}).

\bibitem[{\citenamefont{Siefert and Peinke}(2006)}]{siefert2006}
\bibinfo{author}{\bibfnamefont{M.}~\bibnamefont{Siefert}} \bibnamefont{and}
  \bibinfo{author}{\bibfnamefont{J.}~\bibnamefont{Peinke}},
  \bibinfo{journal}{Journal of Turbulence} \textbf{\bibinfo{volume}{7}},
  \bibinfo{pages}{1} (\bibinfo{year}{2006}).

\bibitem[{\citenamefont{Karth and Peinke}(2003)}]{karth2003}
\bibinfo{author}{\bibfnamefont{M.}~\bibnamefont{Karth}} \bibnamefont{and}
  \bibinfo{author}{\bibfnamefont{J.}~\bibnamefont{Peinke}},
  \bibinfo{journal}{Complexity} \textbf{\bibinfo{volume}{8}},
  \bibinfo{pages}{34} (\bibinfo{year}{2003}).

\bibitem[{\citenamefont{L\"uck et~al.}(2006)\citenamefont{L\"uck, Renner,
  Peinke, and Friedrich}}]{lueck2006}
\bibinfo{author}{\bibfnamefont{S.}~\bibnamefont{L\"uck}},
  \bibinfo{author}{\bibfnamefont{C.}~\bibnamefont{Renner}},
  \bibinfo{author}{\bibfnamefont{J.}~\bibnamefont{Peinke}}, \bibnamefont{and}
  \bibinfo{author}{\bibfnamefont{R.}~\bibnamefont{Friedrich}},
  \bibinfo{journal}{Physics Letters A} \textbf{\bibinfo{volume}{359}},
  \bibinfo{pages}{335} (\bibinfo{year}{2006}).

\bibitem[{\citenamefont{Marcq and Naert}(2001)}]{marcq01}
\bibinfo{author}{\bibfnamefont{P.}~\bibnamefont{Marcq}} \bibnamefont{and}
  \bibinfo{author}{\bibfnamefont{A.}~\bibnamefont{Naert}},
  \bibinfo{journal}{Phys. Fluids} \textbf{\bibinfo{volume}{13}},
  \bibinfo{pages}{2590} (\bibinfo{year}{2001}).

\bibitem[{\citenamefont{Renner et~al.}(2002)\citenamefont{Renner, Peinke,
  Friedrich, Chanal, and Chabaud}}]{renner02a}
\bibinfo{author}{\bibfnamefont{C.}~\bibnamefont{Renner}},
  \bibinfo{author}{\bibfnamefont{J.}~\bibnamefont{Peinke}},
  \bibinfo{author}{\bibfnamefont{R.}~\bibnamefont{Friedrich}},
  \bibinfo{author}{\bibfnamefont{O.}~\bibnamefont{Chanal}}, \bibnamefont{and}
  \bibinfo{author}{\bibfnamefont{B.}~\bibnamefont{Chabaud}},
  \bibinfo{journal}{Phys. Rev. Lett.} \textbf{\bibinfo{volume}{89}},
  \bibinfo{pages}{124502} (\bibinfo{year}{2002}).

\bibitem[{\citenamefont{Friedrich and
  Peinke}(1997{\natexlab{b}})}]{friedrich97}
\bibinfo{author}{\bibfnamefont{R.}~\bibnamefont{Friedrich}} \bibnamefont{and}
  \bibinfo{author}{\bibfnamefont{J.}~\bibnamefont{Peinke}},
  \bibinfo{journal}{Physica D} \textbf{\bibinfo{volume}{102}},
  \bibinfo{pages}{147} (\bibinfo{year}{1997}{\natexlab{b}}).

\bibitem[{\citenamefont{Risken}(1996)}]{risken96}
\bibinfo{author}{\bibfnamefont{H.}~\bibnamefont{Risken}},
  \emph{\bibinfo{title}{The {F}okker-{P}lanck {E}quation}}
  (\bibinfo{publisher}{Springer Verlag}, \bibinfo{year}{1996}).

\bibitem[{\citenamefont{Gardiner}(1985)}]{gardiner85}
\bibinfo{author}{\bibfnamefont{C.~W.} \bibnamefont{Gardiner}},
  \emph{\bibinfo{title}{Handbook of Stochastic Methods}}
  (\bibinfo{publisher}{Springer Verlag}, \bibinfo{year}{1985}).

\bibitem[{\citenamefont{Nocedal and Wright}(1999)}]{nocedal999}
\bibinfo{author}{\bibfnamefont{J.}~\bibnamefont{Nocedal}} \bibnamefont{and}
  \bibinfo{author}{\bibfnamefont{S.~J.} \bibnamefont{Wright}},
  \emph{\bibinfo{title}{Numerical Optimization}}
  (\bibinfo{publisher}{Springer}, \bibinfo{year}{1999}).

\bibitem[{\citenamefont{Byrd et~al.}(1995)\citenamefont{Byrd, Lu, Nocedal, and
  Zhu}}]{byrd1995}
\bibinfo{author}{\bibfnamefont{R.~H.} \bibnamefont{Byrd}},
  \bibinfo{author}{\bibfnamefont{P.}~\bibnamefont{Lu}},
  \bibinfo{author}{\bibfnamefont{J.}~\bibnamefont{Nocedal}}, \bibnamefont{and}
  \bibinfo{author}{\bibfnamefont{C.}~\bibnamefont{Zhu}}, \bibinfo{journal}{SIAM
  J. Sci. Comput.} \textbf{\bibinfo{volume}{16}}, \bibinfo{pages}{1190}
  (\bibinfo{year}{1995}).

\bibitem[{\citenamefont{Zhu et~al.}(1997)\citenamefont{Zhu, Byrd, Lu, and
  Nocedal}}]{zhu1997}
\bibinfo{author}{\bibfnamefont{C.}~\bibnamefont{Zhu}},
  \bibinfo{author}{\bibfnamefont{R.~H.} \bibnamefont{Byrd}},
  \bibinfo{author}{\bibfnamefont{P.}~\bibnamefont{Lu}}, \bibnamefont{and}
  \bibinfo{author}{\bibfnamefont{J.}~\bibnamefont{Nocedal}},
  \bibinfo{journal}{ACM Transactions on Mathematical Software}
  \textbf{\bibinfo{volume}{23}}, \bibinfo{pages}{550} (\bibinfo{year}{1997}).

\bibitem[{\citenamefont{R-2.2.1}(2006)}]{r-project}
\bibinfo{author}{\bibnamefont{R-2.2.1}} (\bibinfo{year}{2006}),
  \urlprefix\url{www.r-project.org}.

\bibitem[{\citenamefont{Chanal et~al.}(2000)\citenamefont{Chanal, Chabaud,
  Castaing, and H\'ebral}}]{chanal2000}
\bibinfo{author}{\bibfnamefont{O.}~\bibnamefont{Chanal}},
  \bibinfo{author}{\bibfnamefont{B.}~\bibnamefont{Chabaud}},
  \bibinfo{author}{\bibfnamefont{B.}~\bibnamefont{Castaing}}, \bibnamefont{and}
  \bibinfo{author}{\bibfnamefont{B.}~\bibnamefont{H\'ebral}},
  \bibinfo{journal}{European Pysical Journal B} \textbf{\bibinfo{volume}{17}},
  \bibinfo{pages}{309} (\bibinfo{year}{2000}).

\bibitem[{\citenamefont{Mor\'e and Thuente}(1994)}]{more1994}
\bibinfo{author}{\bibfnamefont{J.~J.} \bibnamefont{Mor\'e}} \bibnamefont{and}
  \bibinfo{author}{\bibfnamefont{D.~J.} \bibnamefont{Thuente}},
  \bibinfo{journal}{ACM Transactions on Mathematical Software}
  \textbf{\bibinfo{volume}{20}}, \bibinfo{pages}{286} (\bibinfo{year}{1994}).

\end{thebibliography}

\end{document}